\newcommand{\beq}{\begin{equation}}
\def\seq{\end{equation}}
\def\beqs{\begin{eqnarray}}
\def\seqs{\end{eqnarray}}

\def\vf{v_F}
\def\w{\omega}

\def\vf{v_f}
\def\w{\omega}
\def\De{\Delta}
\def\kp{k_\perp}
\def\sgn{{\rm sgn}}
\def\Dn{\Delta  (s, \w_n)}

\def\D2n{\Delta^2(s, \w_n)}

\documentclass[aps,prl,twocolumn,superscriptaddress,floatfix]{revtex4}
\begin{document}

\title{A proposal to  determine the spectrum of pairing-glue in
high-temperature superconductors}
\author{I.~Vekhter}
\affiliation{Theoretical Division, MS B262, Los Alamos National
Laboratory, Los Alamos, NM 87545}
\author{C.~M.~Varma}
\affiliation{Bell Laboratories, Lucent Technologies, Murray Hill,
NJ 07974}

\begin{abstract}
We propose a method for an analysis of the angle-resolved
photoemission data in two-dimensional anisotropic superconductors
which directly yields the spectral function of the bosons
mediating Cooper pairing. The method includes a self-consistency
check for the validity of the approximations made in the analysis.
We explicitly describe the experimental data needed for
implementing the proposed procedure.
\end{abstract}
\maketitle

\paragraph{Introduction.}
Understanding the mechanism of superconductivity in the high
temperature superconducting cuprates requires knowledge of the
spectral function of the bosons mediating the pairing as well as
their coupling to the fermions. For conventional superconductors,
such as Pb or Al, the phonon mechanism of superconductivity was
firmly established through development of a procedure to invert
the tunneling spectra (conductance as a function of voltage) and
obtain the phonon spectral density \cite{Scalapino,Rowell}. Here
we propose a generalization of this procedure for anisotropic
superconductors, which in general may have pairing mediated by the
electron-electron interactions, using the angle resolved
photoemission (ARPES) data as an input.

Recall that in conventional superconductors the dependence of the
normalized tunneling conductance, $G_s(V)/G_n(V)$, where  $s$ and
$n$ refer to the superconducting and the normal state
respectively, on the energy dependent gap function, $\De (\w)$, is
known. In turn, $\De (\w)$ is related to the spectrum of pairing
bosons through the Eliashberg theory \cite{Eliashberg}. This makes
the inversion of the tunneling spectra possible.

Several important differences arise in the case of unconventional
superconductors. First, since for an anisotropic gap function the
tunneling current strongly depends on direction, and since the
tunneling spectra are integrals over all the directions (often
with a varying weight due to the spatial structure of the matrix
element), such spectra in general neither provide sufficient
information to resolve the anisotropy of the boson-fermion
coupling nor allow its frequency dependence to be determined
quantitatively. Second, in contrast to the case of phonons, in
superconductors where the pairing is mediated by electron-electron
interaction, the pairing spectra change and are temperature
dependent below the transition temperature, $T_c$, when the single
particle spectra are gapped. Also, since the typical energy of
interaction between electrons is high, any inversion procedure
must determine whether Eliashberg theory is suitable for the
analysis.

We propose an inversion procedure using the ARPES data, which is
applicable to anisotropic superconductors, uses the Eliashberg
theory and contains a consistency check \cite{Arnold}. Below, we
first summarize the procedure and the experimental input required
to implement it, and then describe the technical details.

\paragraph{Outline of the procedure.}
As input the following experimental information for an optimally
doped or overdoped sample is required: {\sl (i) For
Normalization.} Just above $T_c$, for several points at the Fermi
surface, we require the integrated area under the momentum
distribution curve (MDC) for momentum perpendicular to the Fermi
surface, as defined in Eq.(\ref{IntensityIntegrated}) below. This
step gives the anisotropic equivalent of normal state conductance
in the McMillan-Rowell procedure and provides the necessary
normalization for the second step.
  {\sl (ii) For Inversion.} With
the same polarization and incoming photon energy, and at the same
set of points at the Fermi surface, we need, at several
temperatures below $T_c$, the difference in the intensity of the
MDCs between the normal state above $T_c$ and the superconducting
state, for energies up to several times the superconducting gap.
These data may be either a function of momentum or (better still)
integrated during measurement over the momentum normal to the
Fermi surface as in {\sl (i)}. This step allows the extraction of
the spectral density of the boson and the form of the coupling
function. The data are used to extract the superconducting gap
function, according to Eq.(\ref{ArpesGap}). The gap function is
then inverted by using Eliashberg theory to obtain the boson
spectrum and its coupling to fermions in different angular
channels according to Eq.(\ref{BasicEquation}). The
self-consistency check on the Eliashberg theory is obtained by
comparing the energy range of the bosons with the input ARPES
spectra and with the estimates of the bandwidth.

\paragraph{ARPES spectra.}

The intensity of the ARPES signal, $I({\bf k}, \omega)$, for a
fixed momentum and binding energy of the emitted electron, is
\cite{Randeria}
\begin{equation}\label{ArpesSignal}
I({\bf k}, \omega)={\cal M}({\bf k}) f(\w/T) A ({\bf k}, \omega).
\end{equation}
Here $f$ is the Fermi function, and ${\cal M}$ is a prefactor,
which depends on the square of the dipole matrix element between
the initial and the final states, and on the kinematical factors.
The information about the fermion spectrum
is contained in the spectral function
\begin{equation} \label{SpectralDensityDef}
    A( {\bf k}, \omega) = -\frac{1}{\pi}Im \
        G^R ({\bf k},\omega),
\end{equation}
where $G^R$ is a retarded Green's function. It is the {\bf
k}-dependent prefactor, ${\cal M}$, that obscures the true
momentum dependence of the spectral function, and prevents direct
inversion of the ARPES intensity.

In disentangling the momentum dependence of the spectral density,
$A( {\bf k}, \w)$, from the signal intensity measured in ARPES, we
rely on several properties of the prefactor. First, while ${\cal
M}$ depends on the momentum, {\bf k}, it does not contain a
significant dependence on the temperature, $T$, or the electron
binding energy, $\w$ \cite{Randeria}. Second, since the momentum
dependence of ${\cal M}({\bf k})$ is determined by the spatial
structure of the electron wave functions, the energy and momentum
range over which it varies are of the order of the Fermi energy
and the Brillouin Zone respectively. Consequently, if, for fixed
$\omega$, the spectral density, $A$, is sharply peaked in the
momentum space, the dependence of the prefactor on $|{\bf k}|$ may
be ignored.

\paragraph{Step {\sl (i)}: MDCs and normalization.}
 We write the momentum ${\bf k}=(k_f+k_\perp, s)$,
where $k_f (s)$ is the Fermi momentum and $s$ is the tangential
component of {\bf k} at a point on the Fermi surface. The normal
state Green's function at Matsubara frequencies, $\w_n=2\pi T
(n+1/2)$, is
\begin{equation}\label{GFN}
G_N(\w_n, {\bf k})=\left[
    {i\w_n {\cal Z}({\bf k},\w_n )-\xi({\bf k})}\right]^{-1},
\end{equation}
where ${\cal Z}$ contains the effects of the interactions,
$\xi({\bf k})\approx \vf (s) \kp$, and $\vf$ is the bare,
unrenormalized, Fermi velocity. Therefore the spectral function is
\begin{equation}\label{NSpectral}
    A_N( {\bf k}, \omega) = \frac{1}{\pi}
    \frac{\w Z^{\prime\prime}({\bf k}, \w)}
    {(\w Z^\prime({\bf k}, \w) -\vf\kp)^2+(\w Z^{\prime\prime}(
    {\bf k}, \w))^2},
\end{equation}
where $Z=Z^\prime + i Z^{\prime\prime}$ is the analytic
continuation of ${\cal Z}({\bf k}, \w_n)$ to real frequencies.

Our subsequent analysis assumes that Migdal's theorem is valid for
high-T$_c$ superconductors, and checks for the validity of this
assumption at the end. The essential statement of the theorem is
that, when the energy range of the interaction is small compared
to the electron bandwidth but the momentum ranges of the electrons
and the boson are comparable, (a) the electronic self-energy is
momentum-independent, (b) the vertex corrections to the
interactions are negligible \cite{Migdal}. Under such conditions,
we can approximate $Z({\bf k},\w_n)\approx Z(s, \w_n; k_f)$ in
Eq.(\ref{NSpectral}) to a high degree of accuracy. We believe that
this assumption will be borne out by the self-consistency check
since both the neutron scattering data \cite{Neutrons} and the
analysis of the ARPES data in the normal state
\cite{Valla,Abrahams} indicate the range of the interactions to be
0.2-0.4 eV, only a fraction of the bandwidth.

In contrast to the usual Eliashberg theory, we must consider the
change in the spectrum of the boson in the superconducting state.
We show below that the inversion of the ARPES data at different
temperatures below $T_c$ explicitly gives the evolution of the
pairing interaction as the superconducting order develops. This
point is very important because the value of $T_c$ is determined
by the spectrum just above $T_c$, while all the properties below
$T_c$ (for example the temperature dependence of the
superconducting gap) are determined by the self-consistent change
of the pairing spectrum as a function of temperature.

The momentum distribution curves (MDC), are obtained by plotting
the intensity of the ARPES spectra, $I({\bf k}, \omega)$ at a
fixed electron energy $\w$ as a function of ${\bf k}$
\cite{Valla}.  The MDCs needed here are those taken along the
lines normal to the Fermi surface for several direction. These
curves have Lorentzian shape as a function of $\kp$ \cite{Valla},
which confirms (by inspection of Eq.(\ref{NSpectral})) that both
the renormalization function, $Z$, and the prefactor ${\cal M}$
depend only weakly on $\kp$ in the neighborhood of the peak of the
MDC, ${\cal M}(s, k_f+\kp) \approx {\cal M}(s, k_f)\equiv M(s)$.
\cite{Bansil}

Divide the ARPES intensity by the Fermi function to remove
the temperature dependence, and consider the area under thus
rescaled MDC,
\begin{equation}\label{IntensityIntegrated}
{\cal J}(s, \w)=\int d\kp \frac{I ({\bf k}, \w)}{f(\w/T)}\approx
M(s)
    \int d\kp A({\bf k}, \w).
\end{equation}
It follows from the Lorentzian shape of the
spectral function in the normal state that
\begin{equation}\label{NIntegrated}
    {\cal J}_N(s, \w)\approx \frac{M(s)}{\vf(s)}
\end{equation}
is independent of the binding energy, and only depends on the
position on the Fermi surface via the prefactor and the bare Fermi
velocity. Therefore it serves to normalize the MDCs. It also follows that
 the energy and temperature dependence
of the area under the MDC is simply given by the Fermi function.
Note that to recover most of the area under the MDCs in the
experiment it is sufficient to carry out the integration over the
range of $\kp$ of only a fraction of the Brillouin
Zone.\cite{Valla}

\paragraph{Step {\sl (ii)}: Extraction of the gap function.}
Next consider the difference between the electron Green's
functions in the normal and the superconducting state, and employ
the spectral representation to write this difference as
\begin{eqnarray}\label{GFDiff}
&&G_N({\bf k}, \w_n)-G_s({\bf k}, \w_n)
\\
\nonumber
    &&\qquad =
    \int_{-\infty}^{+\infty} \frac{ dx }{i\omega_n -x}
    \Bigl(A_N({\bf k},x) - A_s({\bf k},x) \Bigr).
\end{eqnarray}
Here $G_s$ is the `11' component of the matrix (Nambu-Gorkov)
Green's function of a superconductor
\begin{equation}\label{GFS}
\widehat G({\bf k}, \w_n)=
    -\frac{i\omega_n {\cal Z}({\bf k},\omega_n) +
        \xi ({\bf k})\tau_3 + \Phi ({\bf k}, \omega_n) (i\tau_2)}
    { {\cal Z}^2({\bf k},\omega_n) \omega_n^2 + \Phi^2({\bf k}, \omega_n) + \xi ^2},
\end{equation}
and $\Phi$ is the anomalous part of the self-energy. We now
integrate both parts of Eq.(\ref{GFDiff}) over the momentum normal
to the Fermi surface at a given point $s$. Using  Migdal's
theorem, we replace both $\Phi$ and ${\cal Z}$ in the integrand by
their values at the Fermi surface, and
 introduce the gap
function $\Delta(s,\w_n)\equiv \Phi(s, \w_n)/Z(s, \w_n)$ to find
\begin{eqnarray}\label{GDiff2}
&&-\frac{i\pi}{\vf(s)}
    \Biggl(\sgn(\w_n)-\frac{\w_n}{\sqrt{\w_n^2 +\D2n}}\Biggr)
\\
\nonumber &&
    \qquad =\int_{-\infty}^{+\infty} \frac{ dx }{i\omega_n -x}
    \int d\kp \Bigl(A_N({\bf k},x) - A_s({\bf k},x) \Bigr).
\end{eqnarray}
Assuming particle-hole symmetry, $A(-k_\perp,s, -x)=A(k_\perp,s,
x)$, and the existence of the center of inversion for the Fermi
surface, ${\cal M}({\bf k})={\cal M}(-{\bf k})$, we rewrite the
integral over the energy, $x$, via an integral over only the
occupied states, $x<0$ (cf.\cite{Norman}). Using the normal state
result, Eq.(\ref{NIntegrated}), to eliminate $M(s)/v_f(s)$, we
find ($\w_n>0$)
\begin{eqnarray}\label{ArpesGap}
    &&\Dn=\w_n \left[\bigl[1-\w_n
    {\cal K}(\w_n,s)\bigr]^{-2}-1\right]^{1/2}
\\
    \label{Kernel}
    &&
    {\cal K}(\w_n,s)=\frac{2}{\pi}\bigl[{\cal J}_N(s)\bigr]^{-1}
\\
    \nonumber
    &&\qquad \times
    \int_{-\infty}^0\frac{dx}{\w_n^2+x^2}
    \bigl[{\cal J}_N(s,x)-{\cal J}_s(s,x)\bigr]
\end{eqnarray}
At each $\w_n=\pi T (2n+1)$ and for given direction $s$ at the
Fermi surface, the function ${\cal K}(\w_n,s)$ depends solely on
experimentally measured intensities. Therefore using
Eqs.(\ref{ArpesGap})-(\ref{Kernel}) the gap function $\Dn $ can be
determined directly from the ARPES spectra.

\paragraph{Inversion of the gap equation.} We assume a pairing
function of the form
    $|{\rm g}({\bf k,k}^\prime)|^2
    B({\bf k}-{\bf k}^\prime, \w_n-\w_{n^\prime})$,
where $|{\rm g}({\bf k,k}^\prime)|$ is the matrix element of the
interaction, and $B({\bf k}, \w_n)$ is the propagator of the
boson. In the Eliashberg theory the equations for the normal and
the anomalous part of the self-energy take the form
\begin{widetext}
\begin{eqnarray}
        \label{EliashbergZ}
    \Biggl[1-{\cal Z}({\bf k}, \w_n)\Bigr] i\w_n=
    -T\sum_{\w_{n^\prime}}\int d{\bf k}^\prime
    |{\rm g}({\bf k,k}^\prime)|^2
    \frac{i\w_{n^\prime} {\cal Z}({\bf k}^\prime, \w_{n^\prime})
          B({\bf k}-{\bf k}^\prime,\w_n-\w_{n^\prime})}
    {\w_{n^\prime}^2{\cal Z}^2({\bf k}^\prime, \w_{n^\prime})+
        \Phi^2({\bf k}^\prime, \w_{n^\prime}) +\xi^2({\bf k}^\prime)
    }
\\      \label{EliashbergF}
    \Phi({\bf k}, \w_n)=\eta
    T\sum_{\w_{n^\prime}}\int d{\bf k}^\prime|{\rm g}({\bf k,k}^\prime)|^2
    \frac{\Phi ({\bf k}^\prime, \w_{n^\prime})
          B({\bf k}-{\bf k}^\prime,\w_n-\w_n^\prime)}
    {\w_{n^\prime}^2{\cal Z}^2({\bf k}^\prime, \w_{n^\prime})+
        \Phi^2({\bf k}^\prime, \w_{n^\prime}) +\xi^2({\bf k}^\prime)
    },
\end{eqnarray}
where  $\eta=+1$ ($\eta=-1$) for coupling in the channel even
(odd) under time-reversal \cite{Sachdev}. Integrating over
$\kp^\prime$ in the right hand side, we arrive at an equation for
the gap function
\begin{equation}
    \label{GapBoson}
    \w_n\Dn=-\pi T\sum_{\w_{n^\prime}}\int ds^\prime
    g(s,s^\prime) B(s-s^\prime, \w_n - \w_{n^\prime})
    \bigl[
            G(s,\w_{n^\prime}) \Dn
        -\eta \w_n F(s,\w_{n^\prime})
    \bigr],
\end{equation}
\end{widetext}
where $g(s,s^\prime)\equiv |{\rm g}(s,s^\prime)|^2/ |\vf (s)|$ is
the effective coupling at the Fermi surface, and $G$ and $F$ are
the $\xi$-integrated diagonal and off-diagonal components of the
Green's function respectively,
\begin{eqnarray}
    \label{Gdef}
    G(s,\w_n)&=&\frac{\w_n}{\sqrt{\w_n^2+\D2n}}
\\
   \label{Fdef}
    F(s,\w_n)&=&\frac{\Dn}{\sqrt{\w_n^2+\D2n}}.
\end{eqnarray}
We now make a simplifying assumption (to be verified) that the
pairing function is separable and diagonal in the space of the
Fermi surface harmonics, $\psi_i(s)$. These harmonics are the
orthonormal eigenfunctions of the symmetry operators of the Fermi
surface, and correspond to different angular channels for a
cylindrical Fermi surface \cite{Comment}. Therefore
\begin{eqnarray}
    \label{VertexExpansion}
   && g(s, s^\prime)B(s-s^\prime, \w_n - \w_{n^\prime})
\\
    \nonumber
    &&\qquad
    =\sum_{i} B_{i} (\w_n - \w_{n^\prime})
    \psi_i(s)\psi_i(s^\prime).
\end{eqnarray}
Expand in $\psi_i(s)$ all functions of $s$ in Eq.(\ref{GapBoson}),
$\Dn=\sum_i \Delta_i(\w_n) \psi_i(s)$ and the same for functions
$G$ and $F$, to obtain a linear system for the coupling function
\begin{equation}
    \label{BasicEquation}
    \sum_{k,\w_{n^\prime}} \beta_{ik}(\w_n, \w_n -\w_{n^\prime})
                                B_k(\w_{n^\prime})=
    \alpha_i (\w_n).
\end{equation}
All the coefficients in this equation are known from the form of
the gap function and the Fermi surface,
\begin{eqnarray}
 \nonumber
   &&\beta_{ik}(\w_n,\w_{n^\prime})=\pi T
                    \Bigl[ \eta\delta_{ik}\w_n F_k(\w_{n^\prime})
                    -b_{ik}(\w_n)G_k (\w_{n^\prime})
                             \Bigr],
\\
    \nonumber
    && b_{ik} (\w_n)=\sum_j a_{ijk}\Delta_j(\w_n),
    \qquad
    \alpha_i (\w_n)=\w_n\Delta_i(\w_n),
\\
     \nonumber
    && a_{ijk}=\int ds \psi_i(s)\psi_j(s)\psi_k(s).
\end{eqnarray}
Solution of  Eq.(\ref{BasicEquation}) gives the pairing function
at Matsubara frequencies and in particular angular channels. It
can then be analytically continued to real frequencies using
either the Pade approximation \cite{Serene} or by solving the
integral equation obtained by analytic continuation of
Eq.(\ref{BasicEquation}) in analogy to Ref.\onlinecite{Carbotte1}.
The full pairing function, $\sum_i
B_i(\w)\psi_i(s)\psi_i(s^\prime)$ can then also be used to obtain
the self-energies, $Z$ and $\Phi$, using
Eqs.(\ref{EliashbergZ})-(\ref{EliashbergF}).

\paragraph{Self-consistency.}
Two major assumptions have been made in our analysis, and have to
be checked for self-consistency. First, we assumed that Migdal's
theorem holds. If the energy range of the pairing function is
significantly smaller than the bandwidth in all channels $i$, the
procedure is self-consistent. Second, we assumed separable form of
the pairing function, Eq.(\ref{VertexExpansion}). In a realistic
calculation we truncate the number of angular channels, $i$, in
Eq.(\ref{BasicEquation}); if inclusion of an additional channel
does not change appreciably the pairing functions in the other
channels, the approximation is justified. Note that a minimum of
two channels is required: the isotropic channel which alone
determines the self-energy in the normal state if the
Fermi-surface is circular, and the $d_{x^2-y^2}$ channel
responsible for pairing; we expect additional channels to enter
with smaller couplings. The change of spectral strength with
temperatures in these channels, obtained via the outlined
procedure, yields crucial information on the change of
self-energies below $T_c$.

\paragraph{Experimental constraints.}
Most available ARPES spectra are taken on
Bi$_2$Sr$_2$CaCu$_2$O$_{8+\delta}$ samples. Since our analysis
includes only superconducting order and does not account for the
pseudogap, the measurements should be done on an optimally doped
or overdoped sample. In the normal state the width of the peak in
MDCs is a fraction of the Brillouin Zone \cite{Valla}, and we
expect that the data in the superconducting state have to be
collected over the same window in momentum space. Since the area
${\cal J}_N$ in Eq.(\ref{NIntegrated}) does not depend on the
energy $\w$, and since MDC is a Lorentzian with a flat
($\kp$-independent) background at $\w=0$, we suggest that ${\cal
J}_N$ is most easily determined at the Fermi energy. Away from the
Fermi energy the background becomes momentum-dependent; however,
the difference between the signal in the normal and
superconducting state, required in Eq.(\ref{Kernel}) is
insensitive to the background. In that equation we expect the
energy range over which the $\kp$-integrated difference in
intensity is appreciable to be of the order of a few times the
superconducting gap, so that the data should be taken up to
$\w\sim 0.1-0.2$eV.

Such data may be difficult to take near the $(\pi,0)$ points
because of the bilayer splitting, not accounted for in our
one-band approach (generalization to a multi-band case is,
however, straightforward).  Also, near these points the MDCs
corresponding to different Fermi arcs begin to overlap at energies
away from $\w=0$ \cite{Valla}. Consequently, it is more feasible
to measure the spectra close to the nodal directions; we expect
that spectra taken over half of the angular range between the node
and the antinode would contain enough information to carry out the
analysis suggested here. Since the analysis of the spectra is
complicated by the superstructure in Bi-O layer, samples
containing Pb, which suppresses Bi-O modulations, may be more
suitable. Finally, the resolution function of the measurement
apparatus has to be disentangled from the signal intensity; the
separation can be done by an integral transform if the energy
dependence of the resolution function is known. A numerical
procedure for implementing the method has been developed and will
be made available.

We have proposed a method to extract the spectrum of the pairing
function from the ARPES measurements in high temperature
superconductors which makes a minimum of theoretical assumptions,
does not use any prejudicial form of the pairing fluctuation
spectrum as an input, and contains a self-consistency test. In
contrast to the analysis of optical conductivity \cite{Carbotte},
our approach uses single particle spectra which contains more
information and explicitly gives the evolution of the pairing
spectrum with temperature in the superconducting state.
Implementation of this approach through appropriate experiments
promises to yield the answer to one of the central questions in
the high $T_c$ problem.

\paragraph{Acknowledgements.} We have benefitted from
discussions with Ar. Abanov and E. Abrahams. This research was
supported in part by the DOE Grant W-7405-ENG-36. I. V. is
grateful to Aspen Center for Physics for its hospitality during
early stages of this work.

\end{document}